# Smartphone Exergames with Real-Time Markerless Motion Capture: Challenges and Trade-offs

**Mathieu Phosanarack**
LAMIH – UMR-CNRS 8201
Univ. Polytechnique Hauts-de-France
Valenciennes, France
mathieu.phosanarack@uphf.fr

**Laura Wallard**
LAMIH – UMR-CNRS 8201
Univ. Polytechnique Hauts-de-France
Valenciennes, France
laura.wallard@uphf.fr

**Sophie Lepreux**
LAMIH – UMR-CNRS 8201
Univ. Polytechnique Hauts-de-France
Valenciennes, France
sophie.lepreux@uphf.fr

**Christophe Kolski**
LAMIH – UMR-CNRS 8201
Univ. Polytechnique Hauts-de-France
Valenciennes, France
christophe.kolski@uphf.fr

**Eugénie Avril**
LAMIH – UMR-CNRS 8201
Univ. Polytechnique Hauts-de-France
Valenciennes, France
eugenie.avril@uphf.fr

## ABSTRACT

Markerless Motion Capture (MoCap) using smartphone cameras is a promising approach to making exergames more accessible and cost-effective for health and rehabilitation. Unlike traditional systems requiring specialized hardware, recent advancements in AI-powered pose estimation enable movement tracking using only a mobile device. For an upcoming study, a mobile application with real-time exergames including markerless motion capture is being developed. However, implementing such technology introduces key challenges, including balancing accuracy and real-time responsiveness, ensuring proper user interaction. Future research should explore optimizing AI models for real-time performance, integrating adaptive gamification, and refining user-centered design principles. By overcoming these challenges, smartphone-based exergames could become powerful tools for engaging users in physical activity and rehabilitation, extending their benefits to a broader audience.

## 1 Introduction

The integration of Information and Communication Technologies (ICTs) in health and rehabilitation has improved patient engagement, adherence, and monitoring [1, 16]. The rise of mobile health (mHealth) has provided a new way for delivering health interventions, leveraging the widespread adoption of smartphones and their accessibility [14]. MHealth solutions offer cost-effective, scalable, and flexible approaches for rehabilitation, particularly in remote or underserved areas [5].

One promising strategy within mHealth is gamification, which involves the use of game design elements, such as rewards, challenges, and progression systems, in non-game contexts [6]. Gamification has demonstrated the potential to enhance user motivation and engagement in various domains [11][18], including health [10], physical activity and rehabilitation [1]. A specific application of gamification in health is exergaming, a category of video games that require players to perform physical exercises to engage with the game [7][22]. Exergames have shown potential in promoting physical fitness, rehabilitation and general well-being [22]. Traditionally, exergames which track user movements, have relied on specialized hardware, such as Microsoft Kinect, VR headsets, and dedicated motion sensors [15][20]. However, such solutions can be costly and may not be available to all [20].

With the proliferation of smartphones, leveraging their built-in cameras for markerless Motion Capture (MoCap) [12] presents a viable alternative that can make exergames more affordable and widely available.

Our goal is to develop a mobile application that enables users to engage in exergames using real-time markerless pose estimation. This application is part of a broader study exploring how gamification can enhance motivation for physical activity. Within this framework, both step counting and exergames contribute to a cumulative score, reinforcing engagement through a gamified intervention.

However, implementing markerless MoCap in smartphone exergames introduces technical and usability challenges, including model accuracy, real-time processing constraints, and interaction design issues. Section 2 presents the background on markerless motion capture and the context of our application. In Section 3, the introduced challenges and their potential trade-offs are discussed to enhance the feasibility and effectiveness of smartphone-based exergames. Finally, Section 4 concludes on our findings while presenting potential directions for future research and improvements.

## 2. Background

### 2.1 Markerless motion capture

Markerless motion capture is a technique that enables the tracking of human movement without the need for physical markers or wearable sensors [12]. It has been widely applied in fields such as biomechanics [17], sports science [17], and clinical rehabilitation [8][12]. Early markerless MoCap systems used multiple RGB and infrared cameras, such as Microsoft's Kinect, which combined depth sensing with machine learning algorithms to track full-body movements in real-time [4].

More recently, advances in artificial intelligence (AI) and computer vision have enabled the development of pose estimation models that use deep learning techniques to estimate body position from RGB camera [19]. Studies comparing models such as OpenPose, PoseNet, and MoveNet have evaluated their suitability for mobile applications [9]. While these models provide varying levels of accuracy and efficiency, many require significant computational power and processing time, making them less viable for real-time applications on smartphones.

Recent developments in lightweight AI models designed for mobile devices have improved real-time MoCap capabilities. Models such as BlazePose [3] and MobileNetV2 [21] optimize efficiency while maintaining reasonable accuracy. These models are integrated into frameworks such as MediaPipe [2], which allows for real-time pose estimation on mobile devices without requiring external sensors.

### 2.2 A gamified mobile application

In our research, we use markerless motion capture within a broader mobile application designed to enhance motivation for physical activity. To leverage gamification as a means of increasing long-term engagement, we developed a mobile application for an extended study to assess its impact on motivation. Within this application, participants can improve their overall score through two main activities: step counting and flexibility exercises. The flexibility exercises are integrated into real-time exergames, which serve for us two main purposes. First, they ensure that users actively perform the exercises to earn a score. Second, they introduce a gamified approach that enhances engagement and enjoyment.

Figure 1 presents the current prototype of the exergame, developed using Unity and Mediapipe. In the figure, body landmarks appear as small cyan and orange dots connected by white lines. Two white spheres positioned at the user's ankles serve as target zones for accumulating points. Additionally, a green and an orange sphere on the user's wrists indicate movement tracking. When a wrist successfully reaches its designated target, it turns green, confirming correct execution, whereas an orange wrist signifies that the target has not yet been reached, resulting in a lower score. When the sphere is too far, it turns red and no score is earned. The total score is displayed at the top of the screen.

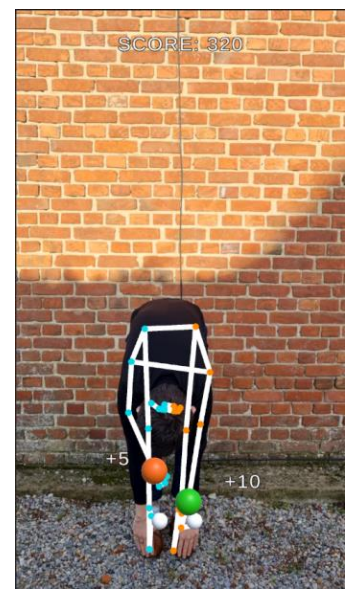

**Figure 1: Prototype of the real-time mobile exergame during flexibility exercise**

During the design and prototyping process, two notable challenges emerged, which may be relevant to researchers, designers, or developers working under similar conditions. The first challenge concerns pose estimation accuracy. In the figure, although the user's hand is correctly positioned, there is a significant offset in the detected hand landmarks, highlighting a limitation in the AI-driven pose estimation model on mobile devices. The second challenge relates to the distance between the user and the smartphone. To ensure full-body pose estimation when the user is standing upright, the user must be positioned relatively far from the device. However, this distance makes it difficult to see the screen clearly, creating an interaction challenge. These challenges are further discussed in the following section.

## 3. Challenges and Trade-Offs

### 3.1 AI model performances

One of the most significant challenges in smartphone-based markerless MoCap for exergames is optimizing computational resources. As presented in Section 2, pose estimation models may require high computational power to function. Moreover, it is important to account for the diverse type of devices which differ from one to another. It is also important to balance accuracy and latency. Indeed, accuracy is critical to validate exercises while low latency is beneficial for responsive gaming experiences.

High-precision AI models often require substantial computational resources, leading to increased processing time and latency [9]. However, optimizing for low latency can compromise accuracy, leading to misinterpretation of user movements and also diminishing gaming experience. Therefore, finding the right balance between accuracy and latency is important to ensure great user experience.

More detailed models, such as OpenPose, provide superior movement tracking but require higher processing power [9], leading to delayed feedback and reduced immersion. For example, OpenPose has a 643 seconds processing time for 1000 images [9] which means that 19.3 seconds are needed to process a one second video at 30 frames per seconds. Even with the fastest model of the study MoveNet Lightning with 53 seconds for 1000 images [9] it needs 1.59 seconds to process a one second video at 30 frames per second.

Therefore, a first potential trade-off is to use lightweight models such as BlazePose and MobileNetV2 to enhance responsiveness [3][21]. Nonetheless, these lighter models may struggle to capture subtle movements or complex poses accurately. Still, depending on the physical exercises, if the range of movement required from user is sufficiently large, this trade-off might be negligible.

Another possible trade-off is considering the usefulness of capturing every movement at each frame. For real-time exergames, there might be no needs in estimating the pose every frame (30 times per seconds in our example). Therefore, limiting the pose estimation to fewer frames could greatly improve the latency. Once again, reducing the number of calculations per seconds could induce diminished capabilities to capture subtle movements. Furthermore, since the pose estimation is less performed, the system responsiveness to user input, which is the body movement, could be impacted. Subsequently, for this trade-off, a new challenge would be to optimize the number of pose estimation performed per second to reduce the latency while maintaining good responsiveness and accuracy.

Moreover, these two solutions can be combined together allowing to reduce latency especially on less powerful devices. However, as previously stated, it is important to preserve a good level of accuracy to not affect user experience.

An additional trade-off to consider is the possibility of off-loading the pose estimation to a server or cloud service. However, this can only be done with very good internet connection and appropriate server calculation power. Indeed, to preserve engaging user interaction, the latency should be kept appropriately low. Work by Liu et al. [13] shows that latency up to 150 milliseconds are found acceptable for multiplayer competitive games. While the stakes of exergames may be lower than competitive game, it highlights the importance of a reliable internet connection for server-based pose estimation.

### 3.2 Interaction challenges

Besides the AI model performances challenges, using markerless MoCap for exergames on mobile device presents specific interaction challenges. One critical challenge is ensuring effective user positioning relative to the smartphone camera. Full-body motion tracking requires users to maintain an appropriate distance from the camera, which may impact their ability to clearly view on-screen instructions. Conversely, standing closer improves screen visibility but may result in incomplete tracking, particularly for lower-body movements.

Potential trade-offs include designing upper-body-focused exergames that prioritize arm and torso movements, allowing users to stay closer to the screen while maintaining accurate tracking. This would reduce the distance between the user and the device, improving the visibility of the smartphone screen. However, the model will not be able to estimate the body lower parts and the artifacts should be thus handled.

Additionally, larger User Interface (UI) elements and alternative feedback mechanisms, such as auditory feedback, can help improve user interactions. Visual instructions could be provided before the exergame, followed by guidance through large and colorful UI elements and audio feedback. Given the variety of feedback modalities, it is essential that they are intuitive and well-explained. One approach is to integrate multiple feedback modalities to ensure that users effectively perceive and interpret the information. Moreover, researchers, designers or developers should consider accessibility for individuals with special needs who may have difficulty relying on auditory, visual or color-based feedback.

An additional interaction challenge to consider is enabling user input outside the immediate exergame context. Due to the real-time markerless MoCap, interacting with the smartphone via touch may not be possible, as users may be positioned too far away. Depending on the gamification approach, the user could fail some exercises, lose points or waste time. To address this limitation, researchers, designers or developers should explore alternative input methods for non-exergame actions. One possible solution is implementing voice or distinct gesture-based commands to pause the game, allowing users to retrieve their smartphone for necessary touch interactions without disrupting gameplay. However, gesture inputs must be carefully designed to prevent unintentional activation during pose-based interactions.

## 4. Conclusion and Future work

This paper outlined key technical and interaction challenges in implementing real-time markerless motion capture for exergames on mobile devices. We discussed potential trade-offs aimed at optimizing performance and user experience. Performance-related trade-offs include balancing accuracy and latency through lightweight AI models, reducing the frequency of pose estimation, and leveraging server-based processing. Interaction-focused trade-offs involve designing upper-body-centric exergames when possible, incorporating larger UI elements which can easily be seen from far away, implementing alternative feedback mechanisms, and exploring diverse user input methods. For our study, these trade-offs will be explored to be implemented to our smartphone-based exergame to improve performances and interaction.

Ongoing and future research, including our work, should focus on enhancing the efficiency and accuracy of pose estimation models on mobile devices to improve real-time performance. Equally important is the need for user studies to assess and refine pose-based interaction methods and feedback mechanisms. In our work and broader research efforts, conducting user evaluations, including questionnaires, can provide valuable insights into the usability, intuitiveness, and effectiveness of these interaction strategies. Such studies would help validate the suitability of different input and feedback approaches, ensuring they enhance user engagement and experience in pose-based exergames.

By addressing these challenges, smartphone-based exergames have the potential to become highly effective, engaging, and widely accessible tools for promoting physical activity and rehabilitation. Markerless MoCap on smartphones offers a cost-effective and scalable solution for integrating exergames into healthcare, enhancing patient engagement, and supporting long-term well-being.